# The effect of Maxwellian fluid on wave propagation in porous media

Weitao Sun[1]*, Fansheng Xiong[1],

[1]Zhou Pei-Yuan Center for Applied Mathematics, Tsinghua University, Beijing, 100084, China.

Corresponding author: Weitao Sun (sunwt@tsinghua.edu.cn)

Running title: **Maxwellian fluid effect on wave propagation**

**Abstract**

This study investigates the effect of a Maxwellian fluid on the propagation of waves in poroelastic media. Based on a fractional derivative stress-strain relation, a viscous dissipation function is obtained to measure the viscoelastic fluid-solid coupling effect. With the viscous dissipation function, elastic waves propagation is formulated in poroelastic media saturated by a fractional derivative Maxwellian fluid, and the analytical expression of the P- and S-wave dispersion/attenuation is presented. Numerical examples show that the fractional derivative Maxwell strain-stress relation has a significant influence on wave velocities and causes the fluid-solid coupling transition from a dissipative regime to an elastic regime. In addition, the predicted fluid velocities are consistent with the laboratory observations of viscoelastic fluids under an oscillating pressure gradient. The results indicate that a viscous-elastic fluid effect may account for the velocity oscillation observed in laboratory. The method elucidates dynamical differences for viscous and viscoelastic fluid in porous medium, which may be of great importance to unconventional oil/gas exploration industry as well as theoretical researches.

# I. Introduction

Elastic wave propagation in poroelastic media is a long-standing problem that has been the focus of numerous papers since the seminal work of Biot(Biot 1956a, b). Many studies have attempted to understand the effect of non-Newtonian fluids on the dynamics of poroelastic media (Pearson and Tardy 2002, Iassonov and Beresnev 2003, Lopez, Valvatne, and Blunt 2003, Harvey and Menzie 1970, De Haro, Del Río, and Whitaker 1996, Mavko 2013, Zhu and Carcione 2014). The nonlinear rheology of the saturating fluid may be a predominant mechanism at low frequencies (Iassonov and Beresnev 2003), and clarifying the role of this mechanism requires a greater understanding of non-Newtonian fluids. Power-law strain-stress relations represent a non-linear model for characterizing aqueous polymer solutions in poroelastic media, such as Bentheim sandstone rock (Teeuw and Hesselink). In addition, the Maxwellian fluid model is another commonly used model for analyzing viscoelastic properties, such as the enhanced oscillating flow rate at specific frequencies (Rio, Haro, and Whitaker 1998). The resonant oscillations has been regarded as a result of non-Newtonian fluid (Tsiklauri and Beresnev 2001, 2003). Experimental observations (Castrejon-Pita et al. 2003) have confirmed the dynamic response of a viscoelastic fluid under an oscillating pressure gradient. Despite the popularity of idealized non-Newtonian models (such as the Maxwell model and the Zener model) (Sochi 2010), the constitutive relationship governing the rheological behavior of a real fluid remains poorly understood, and a more flexible constitutive relationship is required to account for experimental observations.

Maxwellian fluid is usually modeled by a spring-dashpot. However, such a model is usually too specific to describe a real non-Newtonian fluid. Recently, fractional calculus models have received considerable attention because of their ability to characterize viscoelasticity (Friedrich 1991, Heymans 1996, Qi and Liu 2011). Fractional derivative models employ derivatives of a fractional order instead of ordinary derivative operators in the strain-stress relation. The advantage of the fractional constitutive relation relies on its non-local structure, which is suitable for modeling the memory properties of a viscoelastic fluid. Bagley and Torvik (Bagley and Torvik 1983) established a link between fractional calculus models and microscopic theories of ideal viscoelastic media. Butera and Paola (Butera and Paola 2014) showed that a physically based connection occurs between the fractional fluid flowing model and the fractal geometry of a poroelastic medium. A number of recent studies have attempted to analytically account for the flow of a fractional Maxwellian fluid (Tan, Pan, and Xu 2003, Balankin and Elizarraraz 2012, Wang and Tong 2010), and experimental results have confirmed that polymer rheology is consistent with the fractional Maxwell model (Hernández-Jiménez et al. 2002). Transmission of ultrasonic wave in a porous medium has been performed in the frequency domain, which is natural for wave propagation with time harmonic sources (Fellah, Chapelon, et al. 2004, Sebaa et al. 2006, Wu, Xue, and Adler 1990, Santos et al. 1992, Johnson, Plona, and Kojima 1994, Belhocine, Derible, and Franklin 2007). However, direct time domain modeling is more desirable for experimental measurements using pulses of finite bandwidth. Time derivatives of a fractional order has been used to describe the behavior of sound waves in in porous materials (Fellah, Berger, et al. 2004, Fellah and Depollier 2000, Fellah et al. 2003, Fellah et al. 2005). Efforts to model the propagation in the time domain have also been carried out for porous materials with elastic frame (Fellah et al. 2013).

Despite its omnipresence, viscoelastic fluid-solid coupling is still not completely understood by the scientific community. To our knowledge, few studies have focused on the effect of a fractional derivative Maxwellian fluid on elastic wave propagation in poroelastic media, because the viscous and elastic coupling mechanism between a general non-Newtonian fluid and a solid matrix remains poorly understood. In this study, we attempt to formulate the propagation of elastic waves in poroelastic media saturated by a fractional derivative Maxwellian fluid. The attractive feature of fractional derivative method is the reduction of the number of Maxwell elements that have to be accounted for in generalized Maxwell model. In addition, this work explores the effect of rheological parameters and the fractional derivative order on wave velocities by a theoretical analysis and numerical calculations. The results show that the wave velocity dispersion and attenuation of a fractional derivative Maxwellian fluid is quite different from Biot's prediction. Velocity enhancements around a resonant frequency constitute a main feature of the wave field, and the fractional derivative order parameters control the transition from the dissipative regime to the elastic regime.

## II. Frictional dissipation of a fractional derivative Maxwellian fluid

Consider a mass volume $\Omega$ with a closed surface $\partial \Omega$ in a poroelastic medium. Inside volume $\Omega$, the poroelastic matrix is fully saturated by a viscoelastic fluid. The poroelastic matrix is regarded as an isotropic and homogeneous linear elastic body.

The mass continuity equation is

$$\frac{\partial \rho_f}{\partial t} + \nabla \cdot (\rho_f \mathbf{v}_f) = 0 \qquad (1)$$

The momentum conservation equation is

$$\rho_f \frac{D\mathbf{v}_f}{Dt} = -\nabla p + \nabla \cdot \boldsymbol{\tau}, \qquad (2)$$

where $D\mathbf{v}_f/Dt$ is the material derivative, $\mathbf{v}_f$ and $p$ are the fluid velocity and pressure, respectively, and $\rho_f$ is the fluid density. For a fractional derivative Maxwell model, the constitutive equation is (Friedrich 1991, Schiessel et al. 1995)

$$\boldsymbol{\tau} + \lambda^\alpha d_t^\alpha \boldsymbol{\tau} = \eta \lambda^{\beta-1} d_t^{\beta-1} \dot{\boldsymbol{\gamma}}, \qquad (3)$$

where $\boldsymbol{\tau}$ is the extra stress tensor, $\eta$ is the fluid viscosity, $\lambda = \eta/\mu_f$ is the relaxation time and $\mu_f$ is the fluid shear modulus. The deformation rate tensor is defined as $\dot{\boldsymbol{\gamma}} = \nabla \mathbf{v}_f + \nabla \mathbf{v}_f^T - 2/3 \cdot \nabla \cdot \mathbf{v}_f \boldsymbol{I}$. Here the volume viscosity term is omitted because it is important only when the fluid compressibility is essential.

The fractional derivative of order $\alpha$ is defined as (Oldham and Spanier 1974)

$$d_t^\alpha f(t) = \frac{d^\alpha f(t)}{dt^\alpha} = \frac{1}{\Gamma(m-\alpha)} \int_0^t \frac{f^{(m)}(\tau)}{(t-\tau)^{\alpha+m-1}} d\tau, \qquad (4)$$

where $m=1, 2, ...$ and $0 \leq m-1 \leq \alpha \leq m$. To obtain the governing dynamic equation of the fluid phase, take the divergence of both sides of Eq. (3) and substitute it into Eq. (2)

$$\left(1 + \lambda^\alpha D_t^\alpha\right)\rho_f \frac{D\mathbf{v}_f}{Dt} = -\left(1 + \lambda^\alpha D_t^\alpha\right)\nabla p + \eta \lambda^{\beta-1} D_t^{\beta-1}\left(\nabla^2 \mathbf{v}_f + \frac{1}{3}\nabla(\nabla \cdot \mathbf{v}_f)\right). \tag{5}$$

The motion of a fluid in a straight duct of circular cross section provides an ideal case to obtain analytical solution. Here, we formulate an incompressible fractional derivative Maxwellian fluid flowing in a longitudinally oscillating hollow cylinder channel. The long-wave approximation assumes that the pipe radius is much smaller than the wavelength in the pore as well as that of the arrangement of the pores. Therefore, the wall displacement **u** is constant locally. At low speeds, the compressibility of the pore fluid is negligible in relation to the elastic wave propagation, and the mass continuity condition becomes div$\mathbf{v}_f$=0.

For water flowing in a tube with a diameter of serval micrometers, the Reynold number is much less than the critical value. The laminar assumption proves to be reasonable for a viscoelastic fluid flowing in a tube with an even larger radius. At cylinder coordinates $(r, \theta, z)$, we assume that the only nonzero velocity component is along the $z$ (axial) direction and the velocity only depends on the radial coordinate; thus, we have $\mathbf{v}_f^T = \{0, 0, v_{fz}(r)\}$ and $D/Dt = \partial/\partial t$.

In poroelastic media, $\mathbf{v}_f = \mathbf{w} + \dot{\mathbf{u}}$, where **w** is the relative fluid velocity and **u** is the displacement of a solid wall. If the relative fluid velocity $\mathbf{w}(r)$ is independent of the **z** coordinate and axially symmetric, then the governing equation is

$$\rho_f\left(1 + \lambda^\alpha d_t^\alpha\right)\dot{\mathbf{w}} = -\left(1 + \lambda^\alpha d_t^\alpha\right)\left(\nabla p + \rho_f \ddot{\mathbf{u}}\right) + \eta \lambda^{\beta-1} d_t^{\beta-1}\left(\nabla^2 \mathbf{w} + \nabla^2 \dot{\mathbf{u}}\right). \tag{6}$$

Here, the solid-phase velocity $\dot{\mathbf{u}}(z)$ depends on the axial coordinate at the tube surface. The Laplacian of the velocities at the cylinder coordinates is $\nabla^2 \mathbf{w} = 1/r \cdot \partial(r \partial \mathbf{w}/\partial r)/\partial r$ and $\nabla^2 \dot{\mathbf{u}} = \partial^2 \dot{\mathbf{u}}/\partial z^2$. If we use the estimates $\partial^2 \mathbf{w}/\partial r^2 = \mathbf{O}(\mathbf{w}/r^2)$ and $\partial^2 \dot{\mathbf{u}}/\partial r^2 = \mathbf{O}(\dot{\mathbf{u}}/L^2)$, where $L$ is the distance (wavelength) over which significant changes occur, then the long-wave assumption ($r/L \ll 1$) leads to $\partial^2 \mathbf{w}/\partial r^2 \gg \partial^2 \dot{\mathbf{u}}/\partial r^2$.

Assuming that the velocity, displacement and pressure vary in time as $e^{-i\omega t}$, the governing equation is written as

$$\frac{\partial^2 \mathbf{w}}{\partial r^2} + \frac{1}{r}\frac{\partial \mathbf{w}}{\partial r} + \frac{i\omega A}{v}\cdot \mathbf{w} = -\frac{A}{v}\mathbf{F}, \tag{7}$$

where $v=\eta/\rho_f$ and $\mathbf{F} = -(\nabla p + \rho_f \ddot{\mathbf{u}})/\rho_f$. Applying a no-slip boundary condition $\mathbf{w}|_{r=a}=0$ at the fluid-solid wall interface, the solution is

$$\mathbf{w} = -A[1 - J_0(lr)/J_1(la)]\mathbf{F}/l^2 v, \tag{8}$$

where $J_0$ is the Bessel function and $a$ is the radius of the channel; in addition, $l = \sqrt{i\omega A/v}$, where $A = (-iD_e\chi^2)^{1-\beta} + (-iD_e\chi^2)^{\alpha-\beta+1}$ and $D_e = v\lambda/a^2$ is the Deborah number (Reiner 1964, Rio, Haro, and Whitaker 1998).

The average fluid velocity in the cylinder channel can be obtained by integrating $\mathbf{w}$ in cross section:

$$<\mathbf{w}> = \frac{2}{a^2}\int_0^a r\mathbf{w}dr = \kappa(\omega)\mathbf{F}, \tag{9}$$

where

$$\kappa(\omega) = -\frac{A}{l^2 v}\left[1 - \frac{2J_1(la)}{laJ_0(la)}\right]. \tag{10}$$

Here $\kappa(\omega)$ is the dynamic permeability of cylinder channel. $a$ is radius of the channel.

The stress at the fluid-solid wall interface is

$$\boldsymbol{\tau}|_{r=a} = \frac{\eta}{A}\cdot\frac{\partial \mathbf{w}}{\partial r}\bigg|_{r=a} = \frac{\eta}{lv}\cdot\frac{J_1(la)}{J_0(la)}\mathbf{F}. \tag{11}$$

Combining the average fluid velocity $<\mathbf{w}> = 2/a^2 \int_0^a r\mathbf{w}dr$ and total friction force $<\boldsymbol{\tau}> = 2\pi a\boldsymbol{\tau}|_{r=a}$ generates $<\boldsymbol{\tau}> = 8\pi\eta F(\chi)<\mathbf{w}>$. Here, the viscous dissipation function $F(\chi)$ is

$$F(\chi) = -\frac{1}{4}\cdot\frac{\chi\Delta}{-\Delta^2/i}\cdot\frac{J_1(\chi\Delta)}{J_0(\chi\Delta)}\cdot\left(1 - \frac{2J_1(\chi\Delta)}{\chi\Delta J_0(\chi\Delta)}\right), \tag{12}$$

where $\chi = a\sqrt{\omega/\nu}$, $\Delta = \sqrt{-(D_e\chi^2)^{1-\beta}(-i)^{2-\beta} - (D_e\chi^2)^{\alpha-\beta+1}(-i)^{\alpha-\beta+2}}$ and $D_e = \dfrac{a^2}{\nu\lambda}$.

The Deborah number is defined as $D_e=\lambda/\lambda_v$, where $\lambda_v = a^2/\nu$. This number represents the ratio of the relaxation time $\lambda$ and the characteristic time $\lambda_v$ of the viscous effect. The Deborah number indicates if the fluid-solid system is in a dissipative regime or an elastic regime. A material with a lower Deborah number behaves as a Newtonian viscous fluid flow. However, elastic solid behaviors dominate as the Deborah number increases and the material enters a non-Newtonian regime.

Note that the correction function $F$, applied to circular duct flowing by Biot (Biot 1956b), measures the deviation from the Poiseuille flow friction at high frequency. Here the correction function $F(\chi)$ depends on parameters such as fluid viscosity, Deborah number, pore radius and dimensionless frequency $\chi$, etc. The introduction of $F(\chi)$ measures the fluid-solid coupling deviation from the Poiseuille flow friction caused by rheology of fractional derivative Maxwellian fluid. There are no fitting parameters in this function. The method in this work extends Biot's theory to more general cases, including porous medium containing heavy oil and polymer solution.

## III. Elastic wave equations with the effect of a fractional derivative Maxwellian fluid

We incorporate the effect of a fractional derivative Maxwellian fluid by using the viscous term $\eta\phi^2 F(\chi)/\kappa$ in Biot's equations, where $\kappa$ is the permeability. The dilatational and rotational wave equations are obtained by applying a divergence and curl operator to the solid and fluid displacements, i.e., $I_1 = \nabla \cdot \mathbf{u}$, $\xi_1 = \nabla \cdot \mathbf{U}$, $\mathbf{s} = \nabla \times \mathbf{u}$, $\mathbf{S} = \nabla \times \mathbf{U}$.

$$(A+2\mu_s)\nabla^2 I_1 + Q\nabla^2 \xi_1 = \rho_{11}\frac{\partial^2 I_1}{\partial t^2} + \rho_{12}\frac{\partial^2 \xi_1}{\partial t^2} - \frac{\eta\phi^2 F(\chi)}{\kappa}\frac{\partial}{\partial t}(\xi_1 - I_1), \qquad (13)$$

$$Q\nabla^2 I_1 + R\nabla^2 \xi_1 = \rho_{12}\frac{\partial^2 I_1}{\partial t^2} + \rho_{22}\frac{\partial^2 \xi_1}{\partial t^2} + \frac{\eta\phi^2 F(\chi)}{\kappa}\frac{\partial}{\partial t}(\xi_1 - I_1). \qquad (14)$$

$$\mu_s\nabla^2 \mathbf{s} = \rho_{11}\frac{\partial^2}{\partial t^2}\mathbf{s} + \rho_{12}\frac{\partial^2}{\partial t^2}\mathbf{S} + \frac{\eta\phi^2 F(\chi)}{\kappa}\frac{\partial}{\partial t}(\mathbf{S} - \mathbf{s}), \qquad (15)$$

$$0 = \rho_{12}\frac{\partial^2 \mathbf{s}}{\partial t^2} + \rho_{22}\frac{\partial^2 \mathbf{S}}{\partial t^2} + \frac{\eta\phi^2 F(\chi)}{\kappa}\frac{\partial}{\partial t}(\mathbf{S}-\mathbf{s}), \quad (16)$$

where $\mu_s$ is the shear moduli of the solid phase, $\phi$ is the porosity, and $Q$ and $R$ are the same coefficients presented in Biot's work and determined according to Biot and Willis (Biot and Willis 1957). The density $\rho_{12}$ represents inertial drag between solid and fluid, which can be expressed as $\rho_{12} = (1-\alpha_\infty)\phi\rho_f$ (Berryman 1980, Haire and Langton 1999). Here tortuosity $\alpha_\infty$ is a purely geometric variable. It has been shown that tortuosity can be expressed as a function of porosity $\phi$ (Berryman 1980) $\alpha_\infty = 1 - r(1-1/\phi)$, where $r$ is to be calculated from a microscopic model. In particular, for spheres $r = 1/2$ is used in this work.

According to plane-wave assumption, the wave field can be expressed as $I_1 = C_1 e^{i(\omega t - \mathbf{k}\cdot\mathbf{x})}$, $\xi_1 = C_2 e^{i(\omega t - \mathbf{k}\cdot\mathbf{x})}$, $\mathbf{s} = \mathbf{D}_1 e^{i(\omega t - \mathbf{k}\cdot\mathbf{x})}$ and $\mathbf{S} = \mathbf{D}_2 e^{i(\omega t - \mathbf{k}\cdot\mathbf{x})}$, where $\omega$ is the angular frequency and $\mathbf{k}$ is the vector wavenumber. Then, the dilatational wave equations become

$$\begin{bmatrix} a_{11}\left(\frac{k}{\omega}\right)^2 + b_{11}, & a_{12}\left(\frac{k}{\omega}\right)^2 + b_{12} \\ a_{21}\left(\frac{k}{\omega}\right)^2 + b_{21}, & a_{22}\left(\frac{k}{\omega}\right)^2 + b_{22} \end{bmatrix} \begin{Bmatrix} C_1 \\ C_2 \end{Bmatrix} = 0. \quad (17)$$

The rotational wave equations are written as

$$\begin{bmatrix} \left(a_{11}^s \frac{k^2}{\omega^2} + b_{11}^s\right)\mathbf{I}, & b_{12}^s \mathbf{I} \\ b_{21}^s \mathbf{I}, & \left(a_{22}^s \frac{k^2}{\omega^2} + b_{22}^s\right)\mathbf{I} \end{bmatrix} \begin{Bmatrix} \mathbf{D}_1 \\ \mathbf{D}_2 \end{Bmatrix} = 0. \quad (18)$$

The coefficients $a_{ij}$, $b_{ij}$, $a_{ij}^s$ and $b_{ij}^s$ are given in Appendix. The P- and S-wave velocities are obtained by setting the determinant of the wave equations to zero. Therefore, the analytical expressions of the velocity dispersion and attenuation can be derived directly.

For very high frequencies, the wavelengths become much smaller than the grain size so that scattering will occur. If the frequency is higher than a characteristic frequency, Biot's theory is no longer valid and must be extended by a correcting function (Biot 1956b). The Biot's characteristic frequency has been defined as $f_c^{Biot} = \dfrac{b}{2\pi\rho_f}$, where $b = \phi^2\eta/\kappa$ (Biot 1956a, b). In the case of fractional derivative Maxwellian fluid, we introduce a characteristic frequency $f_c^{Maxwell} = \dfrac{\phi^2\eta}{2\pi\rho_f\kappa^*}$, where $\kappa^* = c_\kappa\kappa$.

$c_\kappa = \dfrac{\kappa(\omega)}{\kappa_{w=0}}$ is the ratio between the dynamic permeability $\kappa(\omega)$ and the static permeability $\kappa_{\omega=0}$ of a tube. $\kappa_{\omega=0}$ can be expressed by Hagen-Poiseuille equation $\kappa_{\omega=0} = \dfrac{a^2}{8\nu}$. It is straightforward to derive

$$F(\chi) = \dfrac{a}{4l\nu} \cdot \dfrac{J_1(\chi\Delta)}{J_0(\chi\Delta)} \cdot \dfrac{1}{\kappa(\omega)}$$, from which we obtain $c_\kappa = \dfrac{2J_1(\chi\Delta)}{\chi\Delta J_0(\chi\Delta)F(\chi)}$ and

$$f_c^{Maxwell} = \dfrac{\chi\Delta J_0(\chi\Delta)F(\chi)}{2J_1(\chi\Delta)} f_c^{Biot}.$$ Here $\Delta = \sqrt{-(D_e\chi^2)^{1-\beta}(-i)^{2-\beta} - (D_e\chi^2)^{\alpha-\beta+1}(-i)^{\alpha-\beta+2}}$ and

$\chi = a\sqrt{\omega/\nu}$. $D_e$ is Deborah number, The main difference between $f_c^{Maxwell}$ and $f_c$ is that the permeability is a function of pore structure (tube radius) and rheology parameter (viscosity and Deborah number).

## IV. Results

**A. Effect of rheology parameter and fractional derivative order on frictional dissipation**

Wave dispersion and attenuation arise from frictional dissipation on the pore surface when a viscous fluid flows through a poroelastic medium. Here, we quantitatively calculate the function $F(\chi)$ that characterizes the viscous effect of glycerol and CPyCL/NaSal (cetylpyridinium chloride and sodium salicylate solution). The flow behavior and the viscoelastic properties of the solutions depend on critical shear rates above which the shear thinning and thickening phenomena are observed. Such non-Newtonian behavior is controlled by the Deborah number $D_e$, which is calculated from measurable parameters including the fluid density $\rho_f$, viscosity $\eta$, relaxation time $\lambda$ and fluid flow channel radius $a$. The fluid parameters follow Castrejon-Pita and del Rio, et al. (Castrejon-Pita et al. 2003). For glycerol, we have $\rho_f$=1250 kg/m³ and $\eta$=1 Pa•s. For CPyCL/NaSal (used as the Maxwellian fluid), density and viscosity are $\rho_f$=1050 kg/m³ and $\eta$=60 Pa•s. A channel radius a=1.0 mm is used here to illustrate the effect of glycerol and CPyCL/NaSal on $F(\chi)$.

We plot the real and imaginary parts of the $F(\chi)$ for glycerol and CPyCL/NaSal. The relaxation time for glycerol is set to a tiny value ($10^{-30}$ s) to recover the Newtonian fluid regime, whereas 1.9 s is used to account for the Maxwellian fluid behavior of the CPyCL/NaSal (Castrejon-Pita et al. 2003). The Deborah number is around $8.0 \times 10^{-28}$ for glycerol. We also calculate $F(\chi)$ for different fractional orders.

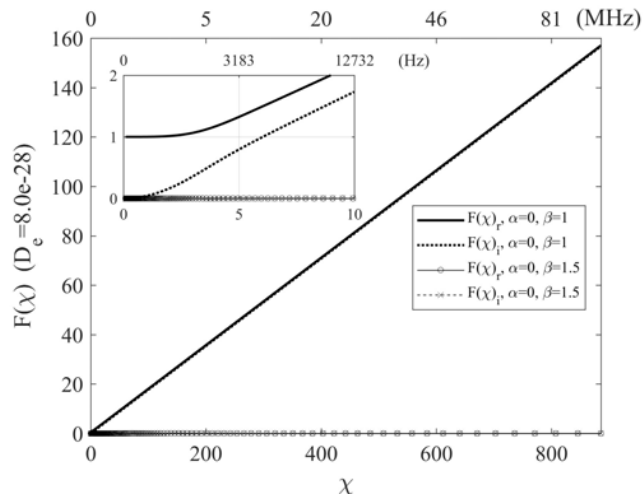

**Figure 1**. Real and imaginary components of the F($\chi$) for glycerol. The fractional derivative order pairs (0, 1.0) and (0, 1.5) are used for comparison. The corresponding frequencies are shown on the top horizontal axis. The embedded subplot shows the details near small $\chi$.

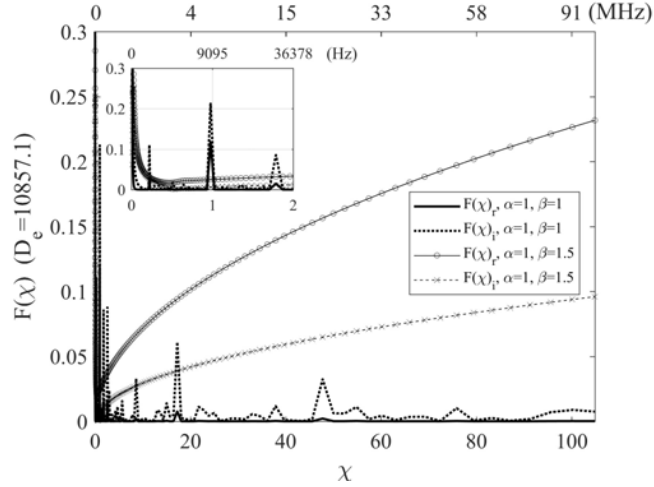

**Figure 2**. Real and imaginary components of the $F(\chi)$ for CPyCL/NaSal. The fractional derivative order pairs (1.0, 1.0) and (1.0, 1.5) are used for comparison. The corresponding frequencies are shown on the top horizontal axis. The embedded subplot shows the details near small $\chi$.

Figure 1 shows that the predicted $F(\chi)$ values of glycerol in this study are exactly the same as the Newtonian limiting case, which was studied by Biot (Biot 1956b) and Tsiklauri (Tsiklauri and Beresnev 2001). This result indicates that a Newtonian fluid can be recovered from the fractional Maxwell model when the relaxation time approaches zero. The fractional derivative order $\beta$ (related to strain retardation) has a significant effect on the $F(\chi)$. A 50% increase in $\beta$ causes $F(\chi)$ to be nearly zeros in Figure 1. The fractional derivative order $\beta$ characterizes the relation between the strain and stress rate for the Newtonian fluid glycerol.

Figure 2 shows the $F(\chi)$ for CPyCL/NaSal. First, $\alpha=1$ and $\beta=1$ are used to recover the conventional Maxwell model. The Deborah number is around 10857.1 for CPyCL/NaSa. Remarkable differences are observed in the frictional dissipation function between glycerol and CPyCL/NaSal. Glycerol has a monotonically increasing $F(\chi)$ with respect to frequency as shown in **Figure 1**, whereas CPyCL/NaSal has a fast decay at low frequency and then presents multiple peaks as shown in **Figure 2**. These peaks represent the resonant response of CPyCL/NaSal fluid. The resonant peaks in the case of Maxwellian fluid has been predicted theoretically and observed experimentally (Castrejon-Pita et al. 2003, Rio, Haro, and Whitaker 1998, Tsiklauri and Beresnev 2001, 2003). Such peaks only occur in viscoelastic fluid cases, rather than in water or brine. Viscous fluid will not show resonance at any frequency.

Then, we apply $α=1$ and $β=1.5$ to illustrate the effect of the fractional derivative orders. The fast decay is still observed at the low frequency regime; however, the multiple peaks are smoothed out as shown in **Figure 2** and $F(χ)$ increases monotonically after the quick decay. The results demonstrate that the fractional derivative orders control the behavior of the resonant peaks. The derivative strain-stress relation has a significant influence on the transition from the dissipative regime to the elastic regime.

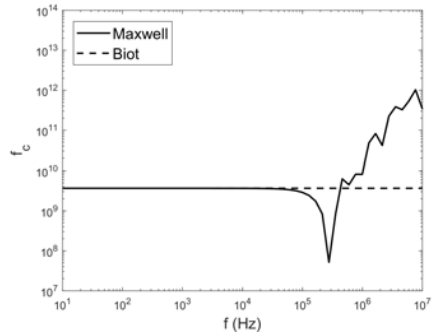

**Figure 3**. Characteristic frequency of Biot's theory and **f**ractional **d**erivative Maxwell (fdMaxwell) model for CPyCL/NaSal at different frequencies. The fractional derivative order pairs (1.0, 1.0) is used for comparison.

The characteristic frequency for CPyCL/NaSal with static permeability of $1.1 \times 10^{-13}$ m² is calculated to show the effect of viscoelastic fluid on wave propagation. The radius $a$ of tube is related to permeability by Hagen-Poiseuille equation $κ_{ω=0} = \dfrac{a^2}{8ν}$, where $ν = η/ρ_f$. The characteristic frequency of Biot's theory is also calculated for comparison. **Figure 3** shows the characteristic frequencies of both Biot's theory and the **f**ractional **d**erivative Maxwell (fdMaxwell) in this work.

One can easily find that Biot's characteristic frequency $f_c^{Biot}$ is a constant value independent on frequency. While for fdMaxwell model, the characteristic frequency $f_c^{Maxwell}$ depends on dynamic permeability $κ(ω)$ and thus is a function of frequency. As we know, for high frequencies, Biot theory is no longer valid because the wavelengths become much smaller than the grain size. Interestingly, we find that $f_c^{Maxwell}$ is the same as $f_c^{Biot}$ as low frequency limit, but increases exponentially at high frequency. The large value of characteristic frequency $f_c^{Maxwell}$ at high frequency clearly demonstrates that fdMaxwell model would be valid even if at extremely high frequency.

**B. Prediction of fluid velocities of glycerol and CpyCl/NaSal in a tube**

The formulation of fluid flow in pore space is an essential part in developing fluid-solid coupling mechanism in poroelastic wave equations. In order to check the theoretically calculated viscoelastic fluid velocity and shear stress in a tube, we will compare the predicted and observed velocities of glycerol and CPyCL/NaSal at the center of the cylinder. The computational parameters follow those in the literature (Castrejon-Pita et al. 2003). For glycerol (used as the Newtonian fluid), $\rho_f$=1250 kg/m$^3$, $\eta$=1 Pa•s, and $\lambda$=0 s, and the fractional derivative orders are $\alpha$=0, $\beta$=1. For CPyCL/NaSal (used as the Maxwellian fluid), $\rho_f$=1050 kg/m$^3$, $\eta$=60 Pa•s, $\lambda$=1.9 s, $\alpha$=1 and $\beta$=1. The cylinder radius is 25 mm, and the displacement amplitude of the oscillation is $z_0$=0.8 mm.

For flow field in a longitudinally oscillating hollow cylinder channel, we use the formulated fluid velocity along the center of the cylinder in Eq. (8). The analytical solution Eq. (8) denotes relative fluid velocity in a poroelastic matrix. Here we assume the wall of cylinder remains still during the experiment. In addition, a low Reynolds number required by Eq. (8) is guaranteed in the experiment (Re < 7x10$^{-4}$). Thus Eq. (8) provides approximated solution of the experimented flow field.

The experimental data of flow velocities of glycerol and CPyCL/NaSal in a cylinder have been reported by Castrejon-Pita and del Rio, et al. (Castrejon-Pita et al. 2003). **Figure 4** shows a sketch of the experimental setup. The transparent cylinder made of acrylic is filled by fluid. The diameter and length of cylinder are d=5 cm and l=50 cm. A piston is driven by a motor in the bottom end of the cylinder, which creates an oscillation at a frequency varying from 1.5 to 200 Hz. The laser Doppler anemometer (LDA) is used to measure local velocity at the center of the cylinder.

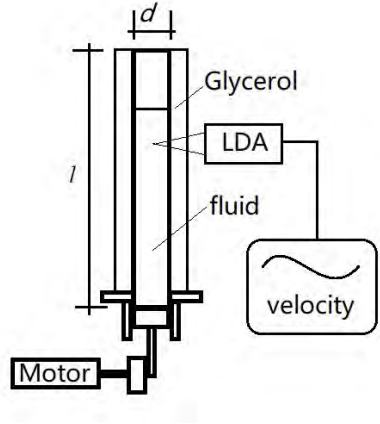

**Figure 4**. Diagram of experiment setup for measuring viscoelastic fluid velocity in a transparent cylinder by laser Doppler anemometer (LDA) probe (following Castrejon-Pita and del Rio, et al., 2003).

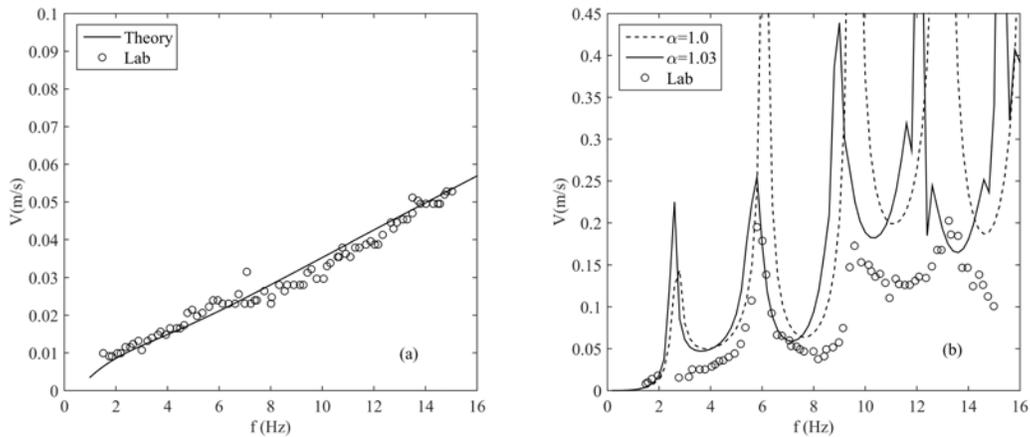

**Figure 5**. Predicted and experimentally measured velocities: (a) glycerol and (b) CPyCL/NaSal. The fractional derivative order is set as 1.0 and 1.03 for CPyCL/NaSal.

Figure 5 shows the significant difference between the glycerol and CPyCL/NaSal fluids. For glycerol, the velocity is nearly a perfect linear function of frequency, whereas for CPyCL/NaSal, the velocity shows multiple peaks, which represent the typical resonance behavior of a Maxwellian fluid. The results confirm that the viscoelastic behavior cannot be ignored when the fluid is driven by an oscillating pressure gradient.

The theoretical predictions in this study predicted the typical viscoelastic resonance behavior as shown in **Figure 5** (b). Although the predicted peak amplitudes are not perfect, the peak locations are consistent. The amplitude discrepancy may arise from the local high shear rate (Castrejon-Pita et al. 2003). Adjusting the fractional derivative order from 1.0 to 1.03 shows that the predicted peaks are much lower and closer to the experimental data. Interestingly, additional small peaks emerge as $\alpha$ increases and the velocity curve becomes less smooth. In addition, the peak positions slightly shift to lower frequencies. The results clearly demonstrate that the fractional derivative Maxwell model is a promising and flexible method to simulate the non-Newtonian fluid effect on wave velocity dispersion.

## C. Effect of fractional derivative order of Maxwellian fluid on P-wave velocity dispersion and attenuation

In this section, we investigate how wave velocities depend on the fractional derivative order of viscoelastic fluid. French Vosgian sandstone with an average porosity of 21% is employed as the poroelastic matrix. The pock parameters (Table 1) have been reported by Bacri and Salin (Bacri and Salin 1986).

Table 1. French Vosgian sandstone properties.

| Property | Value |
| --- | --- |
| $\Phi$ | 0.21 |
| $K_s$ (GPa) | 37 |
| $\rho_s$ (kg/m$^3$) | 2650 |
| $V_p$ (m/s) | 2050 |
| $V_s$ (m/s) | 1240 |
| $\kappa$ (m$^2$) | 1.1x10$^{-13}$ |
| $V_{p\text{brine}}$ (m/s) | 1550 |
| $\rho_{\text{brine}}$ (kg/m³) | 1015 |
| $\eta_{\text{brine}}$ (Pa·s) | 0.001 |

The sandstone is saturated by brine. The relaxation time of the brine is set to a tiny value (10 ns) to simulate the Newtonian fluid behavior. The P-wave velocities of the non-Newtonian wave equation match Biot's theory very well for seismic and ultrasonic frequencies as shown in Figure 6. Tiny differences occur at high frequencies because frictional dissipation is redefined in the wave equations. For slow P-wave in **Figure 7** and S-wave velocities in **Figure 8**, the predicted values for brine-saturation by non-Newtonian wave equations are also the same as those by Biot's theory. The results indicate that for poroelastic media saturated by a Newtonian fluid (brine), which usually has a small viscosity and a linear shear strain-stress relation, Biot's theory works well for modeling elastic wave propagation. In such cases, the viscoelastic effect becomes negligible and the two methods are consistent one with each other.

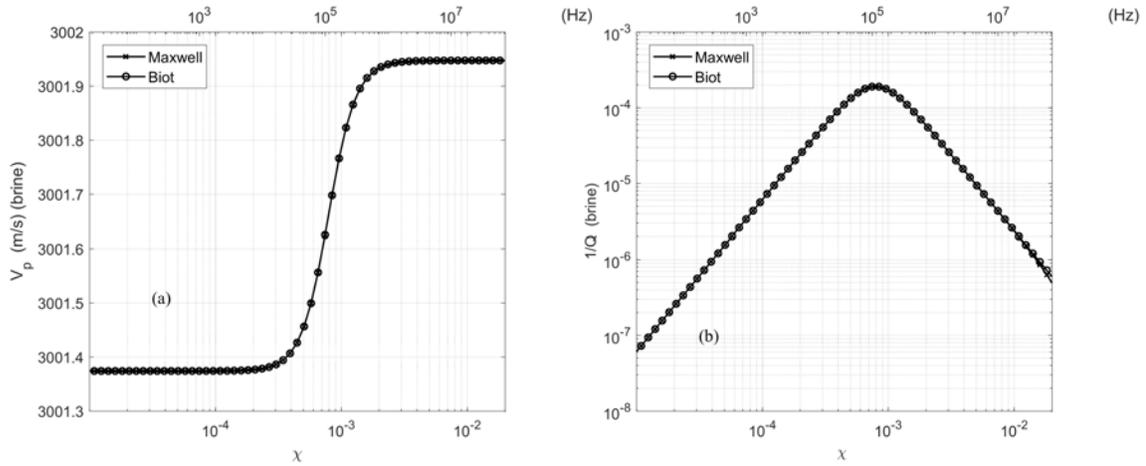

**Figure 6**. P-wave velocity dispersion and attenuation in a poroelastic medium saturated by brine. The velocities are calculated via Biot's theory and the Maxwell model.

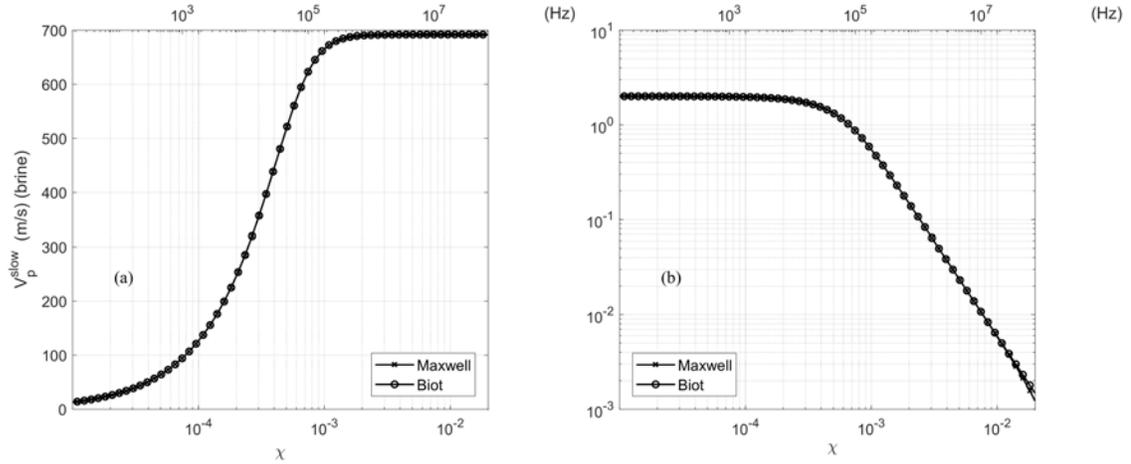

**Figure 7**. Slow P-wave velocity dispersion and attenuation in a poroelastic medium saturated by brine. The velocities are calculated via Biot's theory and the Maxwell model.

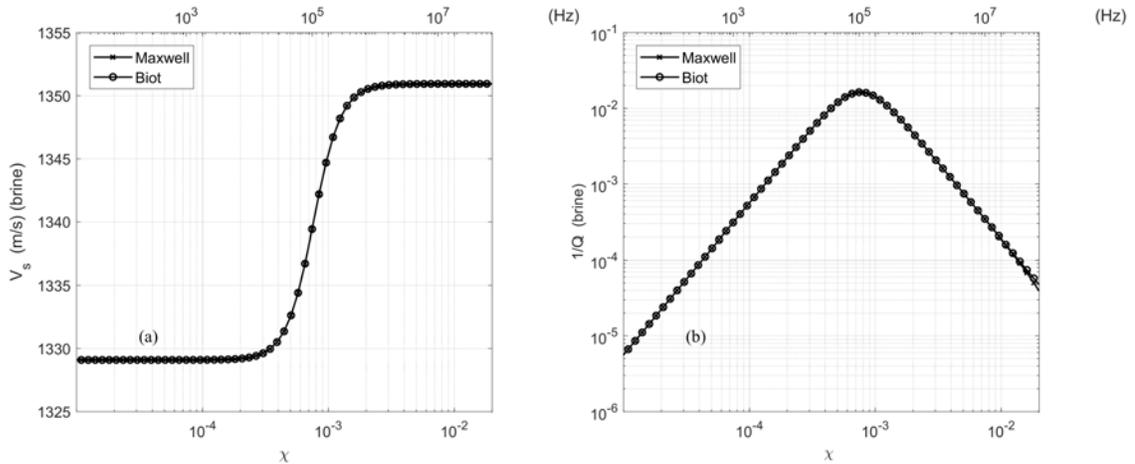

**Figure 8**. S-wave velocity dispersion and attenuation in a poroelastic medium saturated by brine. The velocities are calculated via Biot's theory and the Maxwell model.

Then, we calculate the P-wave velocities for the CPyCl/NaSal saturation as shown in Figure 9. In the case of Biot's theory, the velocity transition and attenuation peak shifts towards a high frequency (from 0.1 MHz in Figure 6 to around 5 GHz in Figure 9) because of the high viscosity of CPyCl/NaSal. However, the peaks of the Maxwell model shift towards a low frequency compared with that under brine saturation. This disparity is closely related to the viscoelastic effect represented by the viscous resistance function $F(\chi)$. The Maxwell model predicts a lower viscous dissipation with sharp fluctuations as shown in **Figure**

**2**, which indicates a higher fluid mobility on average. Moreover, multiple peaks occur in the attenuation curve of the Maxwell model, which likely occurred because the Maxwellian fluid embodies a typical viscoelastic resonance behavior.

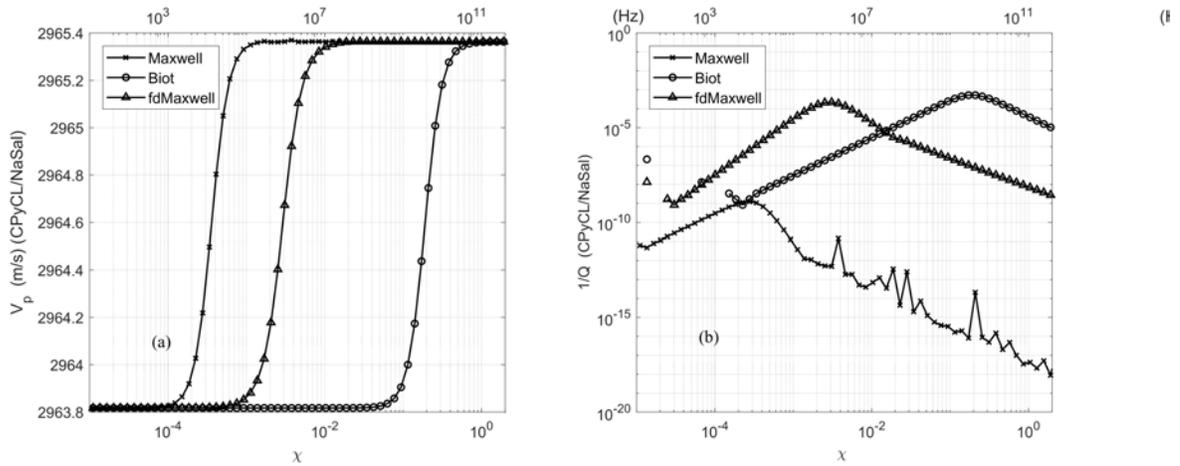

**Figure 9**. P-wave velocity dispersion and attenuation in a poroelastic medium saturated by CPyCl/NaSal. The velocities are predicted by Biot's theory and the fractional derivative Maxwell model with the order parameter (1, 1) (Maxwell) and (1, 1.5) (fdMaxwell).

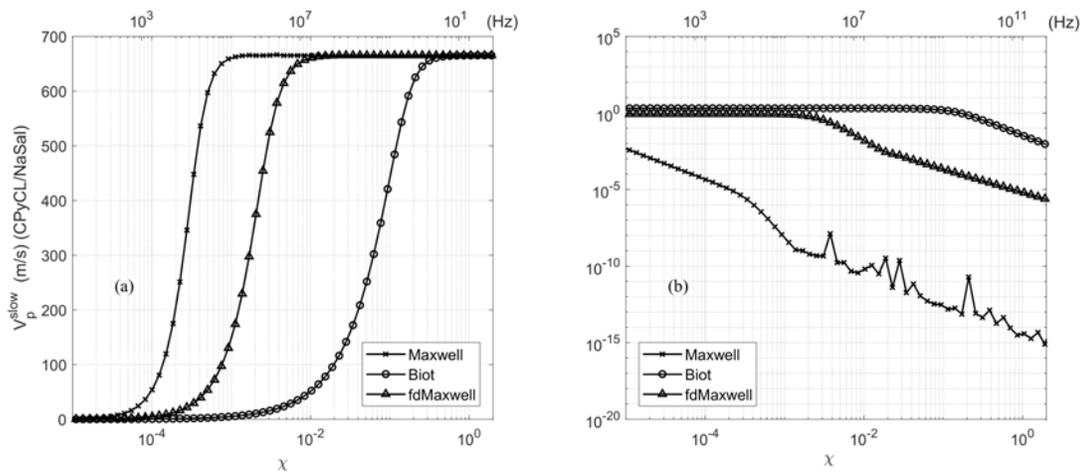

**Figure 10**. Slow P-wave velocity dispersion and attenuation in a poroelastic medium saturated by CPyCl/NaSal. The velocities are predicted by Biot's theory and the fractional derivative Maxwell model with the order parameter (1, 1) (Maxwell) and (1, 1.5) (fdMaxwell).

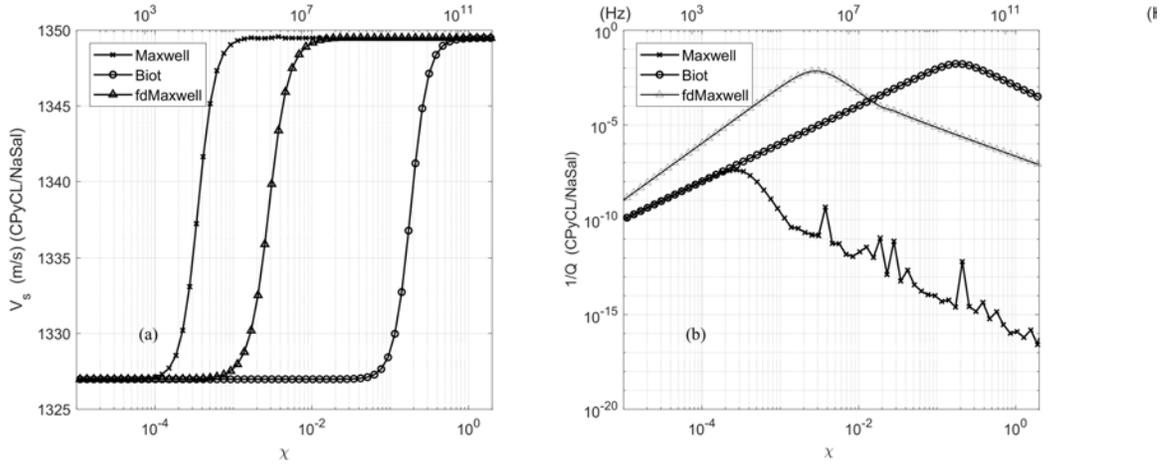

**Figure 11**. S-wave velocity dispersion and attenuation in a poroelastic medium saturated by CPyCl/NaSal. The velocities are predicted by Biot's theory and the fractional derivative Maxwell model with the order parameter (1, 1) (Maxwell) and (1, 1.5) (fdMaxwell).

The predicted slow P-wave velocity in **Figure 10** and S-wave velocity in **Figure 11** again confirm the shift towards low frequency of attenuation peak in Maxwellian fluid case. An interesting observation in slow P-wave velocity is that the attenuation keeps stable from nearly zero frequency to GHz for Biot's theory (**Figure 10** (b)). The velocity dispersion occurs at frequency high above ultrasonic range. However, for wave equations based on fdMaxwell fluid model, the plateau of stable attenuation terminates at a frequency of several MHz. For Maxwellian fluid, **Figure 10** shows that the attenuation decreases over the whole frequency range. A prognosis of the wave equations proposed in this work is that the slow P-wave velocity for viscoelastic fluid has much less attenuation at low frequency, and it is expected that the slow P-wave can be detected much easier in CPyCl/NaSal saturation than in brine saturation.

Compared with P-wave velocity, S-wave velocity has similar dispersion and attenuation behaviors. The velocity transition and attenuation peak position of S-wave are the nearly the same as P-wave in frequency domain, both in the case of brine and in the case of CPyCl/NaSal solution. As expected, the S-wave velocity is much lower than P-wave. However, the S-wave velocity variation range is much larger (about 1329m/s to 1351m/s) than that of P-wave (about 3001.38m/s to 3001.95) in brine saturation case, as shown in **Figure 6** and **Figure 8**. In the case of CPyCl/NaSal saturation, the S-wave velocity variation range is still much larger (about 1327 m/s to 1349 m/s) than that of P-wave velocity (about 2963.8 m/s to 2965.4m/s), as shown in **Figure 9** and **Figure 11**. Amazingly, the slow P-wave has very large velocity variation range in both brine saturation (about 0m/s to 700 m/s) and CPyCl/NaSal saturation (about 0m/s to 670m/s) cases.

Next, $\alpha=1$ and $\beta=1.5$ are employed to investigate the influence of the derivative order on the wave velocities as shown in Figure 9. Interestingly, we find that the dispersion and attenuation curves become much smoother when $\beta$ changes from 1.0 to 1.5. A large value of $\beta$ suppresses the viscoelastic resonance, which was observed in the previous section. In addition, the velocity transition and attenuation peak of the fdMaxwell model shifts to the low frequency region (around 1 MHz) but not as low as that of the Maxwell model (around 10 KHz). The P-wave velocity amplitudes of the fdMaxwell model are at the same level as in Biot's theory, which are quite different from that of the Maxwell model. The order parameters in the fdMaxwell model characterize an intermediate rheological dynamic state between the Maxwell model and Biot's theory. The value of the fractional derivative orders has a remarkable effect on the elastic wave velocities in poroelastic media saturated by a fractional derivative Maxwellian fluid.

Although the fractional derivative order plays an essential role in characterizing the viscoelastic behavior between the purely elastic and the purely viscous, determining the order parameters in the laboratory represents a fundamental challenge. Many efforts have attempted to physically interpret the fractional derivative. For example, the fractal rheological model relates the deformation mechanism of a polymer chain to fractional differential equations. Unfortunately, the relationship between macroscopic stress and microstructural deformation has only been partially resolved for a real polymer liquid, and providing rational fractional derivative parameters to fit a real viscoelastic fluid is difficult. Thus, this topic will not be further examined in this article. Although the current idealized model oversimplifies the molecular-level details, the predicted wave velocities in poroelastic media containing a fractional Maxwellian fluid provide insights on the mechanism underlying the non-Newtonian fluid effect.

## V Conclusions

We formulated the propagation of waves in a poroelastic medium saturated by a fractional derivative Maxwellian fluid. Based on the redefined viscous dissipation function, Biot's theory is extended to account for the effect of a more complex behavior of fluid. The primary importance of this work is the use of wave equations to quantitatively investigate how fluid rheology can be used to determine distinct dynamic responses in poroelastic media. The elastic wave velocity dispersion and attenuation not only depend on the parameters density, elastic moduli and porosity but also on rheological parameters, such as the Deborah number and fractional-derivative orders. Maxwellian fluid causes pulsatile enhancement in wave velocity, which is not observed with a Newtonian fluid. The theoretical results are consistent with the experimental data, and numerical calculations confirm that significant differences occur in the wave velocities in sandstone saturated by Newtonian and non-Newtonian fluids. Our findings indicate that the fractional derivative strain-stress relation controls the transition from a dissipative regime to an elastic regime. The present results clearly demonstrate the importance of incorporating the effect of a fractional derivative viscoelasticity of fluid to understand the mechanism underlying wave propagation in reservoir rocks.

## Acknowledgments

This research study was supported by the National Natural Science Foundation of China (project no. 41874137), the 973 National Key Basic Research Program of China (2014CB239006), the 13th Five-year Basic Research Project of China National Petroleum Corporation (2016A-3301) and the C. C. Lin specific fund.

**Appendix**

The coefficients of the dilatational wave equations are

$$\alpha_{11} = A + 2\mu_s, \qquad b_{11} = -\rho_{11} + i\eta\phi^2 F(c)/\kappa\omega, \qquad \text{A (1)}$$

$$a_{12} = Q, \qquad b_{12} = -\rho_{12} - i\eta\phi^2 F(c)/\kappa\omega, \qquad \text{A (2)}$$

$$a_{21} = Q, \qquad b_{21} = -\rho_{12} - i\eta\phi^2 F(c)/\kappa\omega, \qquad \text{A (3)}$$

$$a_{22} = R, \qquad b_{22} = -\rho_{22} + i\eta\phi^2 F(c)/\kappa\omega. \qquad \text{A (4)}$$

Here, $c = a\sqrt{\omega/\nu}$, $\nu = \eta/\rho_f$, and $b = \eta\phi^2/\kappa$. Thus, the solution of $k/\omega$ is

$$(k/\omega)^2 = \left[-B_p \pm \sqrt{B_p^2 - 4A_p C_p}\right]/2A_p,  \quad \text{A (5)}$$

$$A_p = (A+2N)R - Q^2,  \quad \text{A (6)}$$

$$B_p = -\rho_{11}R - (A+2N)\rho_{22} + 2\rho_{12}Q + ibF(c)(A+2N+2Q+R)/\omega,  \quad \text{A (7)}$$

$$C_p = \rho_{11}\rho_{22} - \rho_{12}^2 - ibF(c)(\rho_{11}+\rho_{22}+2\rho_{12})/\omega.  \quad \text{A (8)}$$

The complex velocity $v_p^* = \omega/k$, phase P-wave velocity $v_p = \text{Re}(v_p^*)$ and attenuation $Q^{-1} = \text{Im}(v_P^{*2})/\text{Re}(v_P^{*2})$ are determined via Eq. A (5).

The coefficients of the rotational wave equations are

$$a_{11}^s = \mu_s, \quad b_{11}^s = -\rho_{11} + i\eta\phi^2 F(c)/\kappa\omega,  \quad \text{A (9)}$$

$$b_{12}^s = -\rho_{12} - i\eta\phi^2 F(c)/\kappa\omega,  \quad \text{A (10)}$$

$$b_{21}^s = -\rho_{12} - i\eta\phi^2 F(c)/\kappa\omega,  \quad \text{A (11)}$$

$$b_{22}^s = -\rho_{22} + i\eta\phi^2 F(c)/\kappa\omega.  \quad \text{A (12)}$$

$$B_s = -\rho_{22}N + iF(c)bN/\omega,  \quad \text{A (13)}$$

$$C_s = \rho_{11}\rho_{22} - \rho_{12}^2 - iF(c)b(\rho_{11}+\rho_{22}+2\rho_{12})/\omega.  \quad \text{A (14)}$$

Thus, the solution of $k/\omega$ is

$$B_s(k/\omega)^2 + C_s = 0.  \quad \text{A (15)}$$

The complex velocity $v_s^* = \omega/k$, phase S-wave velocity $v_s = \text{Re}(v_s^*)$ and attenuation $Q^{-1} = \text{Im}(v_s^{*2})/\text{Re}(v_s^{*2})$ are determined via Eq. A (15).